\documentclass[aps,pra,floatfix,reprint, longbibliography]{revtex4-1}

\usepackage{times}
\usepackage{graphicx}
\usepackage{amsmath}
\usepackage{amssymb}
\usepackage{enumitem}
\usepackage{mathtools}
\usepackage{xcolor}
\newcommand{\Mod}[1]{\ \mathrm{mod}\ #1}

\begin{document}

\title{Quantum-classical hybrid algorithm using an error-mitigating $N$-representability condition to compute the Mott metal-insulator transition}

\author{Scott E. Smart and David A. Mazziotti}
\email{damazz@uchicago.edu}

\affiliation{Department of Chemistry and The James Franck Institute, The University of Chicago, Chicago, IL 60637 USA}

\date{Submitted February 18, 2019; Revised July 30, 2019}

\begin{abstract}
Quantum algorithms for molecular electronic structure have been developed with lower computational scaling than their classical counterparts, but emerging quantum hardware is far from being capable of the coherence, connectivity and gate errors required for their experimental realization. Here we propose a class of quantum-classical hybrid algorithms that compute the energy from a two-electron reduced density matrix (2-RDM). The 2-RDM is constrained by $N$-representability conditions, conditions for representing an $N$-electron wave function, that \textcolor{black}{mitigates} noise \textcolor{black}{from} the quantum circuit. We compute the strongly correlated dissociation of doublet H$_{3}$ into three hydrogen atoms. The hybrid quantum-classical computer matches the energies from full configuration interaction to 0.1~kcal/mol, one-tenth of ``chemical accuracy,'' even in the strongly correlated limit of dissociation. Furthermore, the spatial locality of the computed one-electron RDM reveals that the quantum computer accurately predicts the Mott metal-insulator transition.
\end{abstract}

\maketitle

\section{Introduction}
Quantum computers hold the promise of tackling some of the most challenging simulations of many-electron quantum systems~\cite{Lloyd1996,Aspuru-Guzik2005}. A number of quantum algorithms have been developed which exhibit lower scaling than their classical counterparts~\cite{Abrams1997,Abrams1999,Farhi2000}, but emerging quantum hardware is far from being capable of long coherence times, arbitrary connectivity and low gate error, which are requirements for most of these algorithms. As a consequence, efforts to maximally utilize the available devices have taken inspiration from quantum and classical regimes alike~\cite{Wang2008,Devoret2013,Moll2017,Kandala2018,Bian2018}. In particular, hybrid quantum-classical algorithms have been developed, which attempt to separate efficiently quantum and classical components of a problem~\cite{Peruzzo2014,Kandala2017,Santagati2018}. Quantum hardware is used to prepare and measure a quantum state, or encode information, with the remaining tasks distributed to a conventional computer for classical execution~\cite{Peruzzo2014, McClean2016}.  \textcolor{black}{Attempts to minimize the effect of the noise through algorithm design are known as error mitigation.  Error mitigation schemes, recently proposed and implemented, include those based on extrapolative procedures or inherent stabilizer codes~\cite{Li2017,Temme2017,Kandala2019,Bonet-Monroig2018}.}

While recent hybrid quantum-classical algorithms like the quantum eigensolver method compute the two-electron reduced density matrix (2-RDM) to determine the energy, they are developed with the wave function's variational principle and hence, they do not consider the 2-RDM's variational principle. The key distinction between these two variational approaches is that the variational principle of the 2-RDM contains additional constraints that are necessary for the 2-RDM to represent at least one $N$-electron density matrix or wave function, known as $N$-representability conditions~\textcolor{black}{\cite{Coleman1963,Coleman1978,Mazziotti2007,Mazziotti2012}}.  On a classical computer necessary $N$-representability conditions allow us to compute a lower bound on the ground-state energy and an approximate 2-RDM without computation or storage of the $N$-electron wave function~\cite{Erdahl2000, Nakata2001, Mazziotti2001, Zhao2004, Mazziotti2004, cances2006electronic, mazziotti2006variational, Mazziotti2007, gidofalvi2008active, shenvi2010active, Mazziotti2011, verstichel2012variational, Mazziotti2016a}.  The variational calculation of the 2-RDM subject to approximate $N$-representability conditions can capture strong electron correlation in molecular systems at a computational cost that scales polynomially with the number $N$ of electrons~\cite{greenman2010strong, Schlimgen2016, fosso2016large, Montgomery2018, Safaei2018, Sajjan2018}.  While a perfect quantum computer would be able to operate with the variational principle of the wave function, near-term quantum computers operate with substantial noise that disrupts the $N$-representability of the measured 2-RDM.  The 2-RDM principle provides a physical resource for error \textcolor{black}{mitigation} in the form of the $N$-representability conditions. Previous work has considered the use of these conditions to perform quantum tomography of 1-electron RDMs (1-RDMs) and 2-RDMs from noisy experimental data, and more recent work has proposed the extension of these ideas to measurements from a quantum computer~\cite{Foley2012, Rubin2018, Sagastizabal2019}.

In this paper we propose and implement a quantum-classical hybrid algorithm for molecular electronic structure that uses a 2-RDM variational principle in which the 2-RDM is constrained by $N$-representability conditions.  Previous electronic structure calculations on quantum computers have largely treated 2- or 4-electron atoms or molecules in closed-shell states without significant electron correlation~\cite{Shen2017,McClean2017,Colless2018,Kandala2017, Kandala2018}. We implement an algorithm for 3-electron molecules in open-shell, doublet states with significant strong electron correlation. A pure-state $N$-representability condition, known as a generalized Pauli constraint, which was originally discovered by Borland and Dennis at IBM in a series of computations on a classical computer~\cite{Borland1972}, allows us to express the $N$-representable 2-RDM for three-electron systems as a functional of only the 1-RDM~\cite{Chakraborty2014,Schilling2017}.  We optimize the eigenvalues of the 1-RDM on the quantum computer and its eigenfunctions, which are not restricted by $N$-representability, on the classical computer.  \textcolor{black}{The eigenvalues of the 1-RDM are represented by an 3-electron wave function on the quantum computer.}  Computation of the strongly correlated dissociation of molecular H$_{3}$ yields its potential energy surface and an accurate prediction of its Mott metal-to-insulator transition~\cite{Schuster19}. Energies are computed with errors of about 0.0001 atomic units (or less than 0.1 kcal/mol).

\section{Theory}

\textcolor{black}{After discussing a general quantum-classical hybrid algorithm for computing the ground-state energy and 2-RDM with error-mitigating $N$-representability conditions in section~\ref{sec:nrep}, we examine the details of implementing such an algorithm for 3~electrons in 6~orbitals in sections~\ref{sec:err} to \ref{sec:alg}.}

\subsection{Quantum-classical Hybrid Algorithm with $N$-Representability Conditions}

\label{sec:nrep}

For an $N$-electron system ($N\geq2$), we can write the two-electron reduced density matrix (2-RDM) as: \textcolor{black}{
\begin{equation}
^2 D(12,\bar{1}\bar{2}) = \int \psi(123...N) \psi^*(\bar{1}\bar{2}3...N) d(3...N).
\end{equation}}
The 2-RDM of a system has all the information necessary for calculating the energy and other molecular properties. For instance, the energy of a molecular system is obtained as:
\begin{equation}\label{energy}
E = {\rm Tr} (^2K \ ^2D),
\end{equation}
where $^2K$ is the reduced Hamiltonian. Although the energy is expressible as a linear functional of the 2-RDM, the 2-RDM must be constrained by $N$-representability conditions for it to be representable by at least one $N$-electron density matrix or wave function~\cite{Coleman1978,Mazziotti2007,Mazziotti2012,Mazziotti2016b}. Necessary ensemble-state and pure-state $N$-representability conditions are known~\cite{Mazziotti2012,Altunbulak2008,Mazziotti2016b}.

\textcolor{black}{A general quantum-classical hybrid algorithm for computing the ground-state energy and 2-RDM with error-mitigating $N$-representability conditions is given in Table~\ref{t:QCA} for $N$-electron quantum systems.  After the quantum state is prepared on the quantum computer through a series of unitary transformations, tomography is performed to measure the elements of the 2-RDM.  Unlike traditional algorithms, the 2-RDM in step~3 is corrected for errors from hardware or noise by accounting for additional constraints on the 1- or 2-RDM~\cite{Foley2012,Rubin2018,Sagastizabal2019} such as ensemble or pure-state $N$-representability conditions on the 2-RDM~\cite{Mazziotti2012, Mazziotti2016b}.   For example, in the next section, we discuss using a pure $N$-representability condition for 3 fermions, and more generally, Foley and Mazziotti~\cite{Foley2012} discuss general corrections of RDMs for ensemble $N$-representability conditions through semidefinite programming.  The performance of orbital rotations on the classical computer, as proposed here in step~5, can be applied to any electronic system with any number $N$ of electrons because $N$-representability conditions are invariant to unitary transformations of the orbitals~\cite{Coleman1963}.   Orbital rotations of the 2-RDM can be performed with polynomial-scaling cost on the classical computer, which simplifies the quantum circuit required on the quantum computer, thereby decreasing the effects of hardware errors and noise.  In step~6 the unitary transformations that prepare $\Psi$ are updated via a derivative-free optimization algorithm.  Finally, steps 1-6 are repeated until the ground-state energy converges below a given threshold $\epsilon$.}

\begin{table*}[t!]

\caption{Quantum-classical hybrid algorithm for the ground-state energy and 2-RDM with error-mitigating $N$-representability conditions.}

\label{t:QCA}

\begin{ruledtabular}
\begin{tabular}{l}
{\bf Algorithm: Quantum-classical hybrid algorithm for the 2-RDM with error-mitigating $N$-representability conditions} \\
\hspace{0.5in} {Given a convergence threshold} $\epsilon$. \\
\hspace{0.5in} {Choose the initial unitary transformation to prepare} $\Psi$. \\
\hspace{0.5in} {Repeat until convergence of the ground-state energy.} \\
\hspace{1.0in} {{\bf Step 1:} Prepare} $\Psi$ {via unitary transformations (quantum computer)}, \\
\hspace{1.0in} {{\bf Step 2:} Perform tomography to measure the elements of the 2-RDM (quantum computer)}, \\
\hspace{1.0in} {{\bf Step 3:} Correct 2-RDM for a set of $N$-representability conditions (classical or quantum computer)},\\
\hspace{1.0in} {{\bf Step 4:} Compute the energy from the 2-RDM from} ${\rm Tr}(^{2} K \, ^{2} D)$ {(classical computer)},\\
\hspace{1.0in} {{\bf Step 5:} Minimize the energy with respect to orbital rotations (classical computer)},\\
\hspace{1.0in} {{\bf Step 6:} Update parameters in the unitary transformations from derivative-free optimization (classical computer)},\\
\end{tabular}
\end{ruledtabular}

\end{table*}

The pure-state $N$-representability conditions of the 1-RDM, also known as the generalized Pauli constraints~\cite{Borland1972, Klyachko2006,Altunbulak2008, Schilling2013, Chakraborty2014, Benavides-Riveros2015, Schilling2015, Theophilou2015, Mazziotti2016b, Schilling2017, Chakraborty2018}, are in the form of linear inequalities on the set of 1-RDM eigenvalues (natural occupation numbers) for a given number of electrons and orbitals.  In 1972 Borland and Dennis~\cite{Borland1972} discovered these constraints that extend the Pauli exclusion principle in the case of 3 electrons in 6 orbitals, and in 2006 Klyachko (and in 2008 with Altunbulak) generalized their derivation for potentially arbitrary numbers of electrons and orbitals\cite{Klyachko2006,Altunbulak2008}.  In general these constraints are not saturated by the natural occupation numbers of correlated quantum systems~\cite{Mazziotti2016b}.  However, these constraints are often quasi-saturated (or quasi-pinned)~\cite{Schilling2013}, and in the case of atoms and molecules with 3 electrons in 6 orbitals it has been computationally demonstrated that in many cases the generalized Pauli constraints are saturated by the natural occupation numbers ~\cite{Chakraborty2014, Benavides-Riveros2015}.

In this work we focus on the 3-electron-in-6-orbital system, which has eigenvalues of the 1-RDM, or natural occupation numbers, $\{n_i\}$ (and $1\leq i \leq 6, i\in\mathbb{N}$), where $n_i\geq n_{i+1}$.  The constraints on $\{n_i\}$ in this case, known as the Borland-Dennis constraints~\cite{Borland1972}, are as follows:
\begin{eqnarray}
n_5+n_6-&n_4 \geq 0 \label{bdc} \\
n_1+n_6 &=1 \\
n_2+n_5 &= 1 \\
n_3+n_4 &=1 \label{bdc4}
\end{eqnarray}
When a wave function saturates the inequality, then its expansion contains only Slater determinants that also saturate the inequality, which is known as a selection rule, and hence, in this case only 3 determinants contribute to its expansion~\cite{Benavides-Riveros2016, Chakraborty2018, Boyn2019}:
\begin{equation}\label{gpcwf}
|\psi \rangle = \alpha |111000\rangle + \beta |100110 \rangle + \gamma|010101\rangle
\end{equation}
where $|\gamma|^2 = (1-|\alpha|^2-|\beta|^2)$, $1\geq |\alpha|^2 \geq |\beta|^2$, and $|\alpha|^2 \geq |\beta|^2+|\gamma|^2$. Any two $n_i,n_j$ of the 1-RDM where $i+j \neq 7$, are linearly dependent basis vectors of the wavefunction $| \psi \rangle$ up to phases. As discussed in the paragraph above, this saturation has been demonstrated computationally for the ground states of a wide variety of 3-electron-in-6-orbital atoms and molecules~\cite{Chakraborty2014,Benavides-Riveros2015}.  The  3-electron-in-6-orbital atoms and molecules need not be pinned to the Borland-Dennis inequality~\cite{Schilling2013,Chakraborty2014,Schilling2017}, but such pinning has been observed in Li, the potential energy surface of H$_{3}$, the $\pi$ system of C$_{3}$H$_{3}$, as well as other systems~\cite{Chakraborty2014, Chakraborty2018}.  Using this pruned expansion, we are able to carefully reconstruct the 2-RDM from the 1-RDM (see Appendix C).  The reconstruction of the 2-RDM in this case is equivalent to the correction of the 2-RDM by $N$-representability conditions in the general algorithm.

\subsection{Error Mitigation Scheme}

\label{sec:err}

While the $N$-representability condition itself acts like a form of error correction by constraining the 1- and 2-RDMs to be representable by a wave function, in obtaining 1-RDM's from a quantum computer an additional error \textcolor{black}{mitigation scheme} can be implemented to ensure that all permissible 1-RDM eigenvalues are explored. The set of occupation numbers that satisfy the pure constraints on a system forms a multi-dimensional convex set with ``flat'' sides known as a polytope. \textcolor{black}{For instance, in the case of 3-electrons in 6-orbitals, the Pauli exclusion principle defines 4 ``planes'' in the space spanned by $\{ n_4,n_5,n_6\}$:
\begin{eqnarray}
n_6-n_5 & =0 \label{ord1}\\
n_5-n_4 & =0 \label{ord2}\\
n_6 &=0 \label{ferm1}\\
n_4-\frac{1}{2} &= 0, \label{ferm2}
\end{eqnarray}
 where Eqs.~\eqref{ord1} and \eqref{ord2} are from the ordering constraints, and Eqs.~\eqref{ferm1} and \eqref{ferm2} are from Pauli-exclusion limits on occupations. The generalized Pauli constraints provides another plane, defined as Eq.~\eqref{bdc} (which actually is stronger than Eq.~\eqref{ferm1}). The intersection of these planes provides the relevant polytope.} The Hartree-Fock (or initialized qubit) state is one extrema of \textcolor{black}{the} polytope with the constraints defining the facets of the polytope\textcolor{black}{, given as $v_{HF}=(0,0,0)$.}

The basic principle of the error \textcolor{black}{mitigation} is to remap the extrema of the polytope to adjust for errors detected through an initial set of measurements. In other words, let the set of all points in the polytope under the pure constraints be $A$, and the set of measurable points under error be $A'$. Then, we introduce a mapping $T$:
\begin{equation}\label{ec}
T: A' \rightarrow A.
\end{equation}
and $A$ includes the desired region. In the present work we employ this as a simple affine transformation.

\subsection{Hybrid Variational Algorithm}

\label{sec:alg}

The optimization of the 1-RDM is carried out in the natural-orbital basis set on the quantum computer. Orbital rotations, which are necessary to determine the energetically optimal natural orbitals, scale polynomially with the number of orbitals and hence, are treated on the classical computer. This partitioning of tasks between the quantum and classical computers, physically motivated by the structure of the $N$-representability conditions, reduces the complexity of the optimization.

To stay in the natural orbital basis, we use the following 3-qubit gate sequence:
\begin{equation}\label{qalg1}
\hat{U}(\theta_1,\theta_2)= C_2^1 R_2^y(\theta_2) C_1^3 R_1^y (\theta_1)
\end{equation}
where $C_i^j$ and $R_j^y(\theta_i)$ are the controlled-NOT and Y-rotation gates. \textcolor{black}{This series of transformations was derived to ensure that the set of resulting states spans the plane of occupations in} \textcolor{black}{Eq.~\eqref{bdc}-\eqref{bdc4}.} The quantum state has a diagonal 1-qubit RDM, and a standard qubit measurement in the computational basis is sufficient to obtain the 1-RDM.  \textcolor{black}{Eqn.~(\ref{qalg1}) is the state preparation which can be replaced with unitary transformations of increasingly complexity for any number $N$ of electrons.  Furthermore, in step~3 of the general algorithm in Table~\ref{t:QCA} the Borland-Dennis constraint can be replaced with a set of more general $N$-representability conditions ~\cite{Mazziotti2012, Mazziotti2016b}, using an RDM-correction scheme such as the one described in Ref.~\onlinecite{Foley2012}.}

The set of possible occupation numbers of a 1-RDM generated by Eq.~\eqref{qalg1} forms a curved two-dimensional surface within the polytope. A transformation $T_i$ from the qubit space to the molecular space of Eq.~\eqref{gpcwf} is given by
\begin{equation}
\begin{split}
T_i = & \  G_i Q_i^{-1}   \\    T_i :\mathbb{C}^{2^{3}} &\rightarrow  \wedge^3\mathcal{H}_6,
\end{split}
\end{equation}
where $G_i$ are vertices of the space spanned by Eq.~\eqref{gpcwf} subject to ordering constraints, and $Q_i$ are vertices of the measured set of the algorithm (see Eq.~\eqref{qalg1}). The $i$ indices correspond with the triangulation of the curved surface. Note that when compared with Eq.~\eqref{ec}, we use a composition to map directly to the desired 3-electron-6-orbital  Hilbert space.

During the optimization, the 1-qubit RDM is measured on the quantum computer, the error \textcolor{black}{mitigating transformation} $T_i$ is applied, and the 2-RDM is constructed classically from the corrected 1-RDM elements.  Following the convergence with respect to $^2D$, Givens rotations are carried out classically on the $^2K$ matrix to minimize the energy according to Eq.~\eqref{energy}. The process is carried out iteratively until both methods converge. The Nelder-Mead simplex or steepest descent algorithms were used for both optimizations. \textcolor{black}{See Appendix A for more details.}

\section{Results and Discussion}
\subsection{Dissociation of Linear H$_3$}

Figure 1 presents the dissociation of the linear H$_3$ molecule in its ground doublet state into three hydrogen atoms from equal stretching of the two bonds. Calculations were performed in the Slater-type-orbital minimal basis set (STO-3G) with results compared to those from classical full configuration interaction (FCI). Using the RDM method on the quantum computer (RDM-QC), we obtain a highly accurate potential energy curve for the H$_{3}$ dissociation even for values of the internuclear distance greater than 2~\AA\ where strong electron correlation is present due to the spin entanglement among the energetically degenerate H-atom 1s orbitals. Traditional single-reference methods like second-order many-body perturbation theory or the coupled cluster singles-double method either diverge or fail to converge as the bond is stretched beyond 2~\AA . Throughout the dissociation curve energy errors from RDM-QC are consistently less than 0.0001 a.u. (or less than 0.1 kcal/mol) relative to FCI.  \textcolor{black}{Because the RDM reconstruction guarantees a physical, $N$-representable 2-RDM, the energy curve obtained is an upper bound to the FCI energy curve.  While uncertainty from sampling on the order of $0.002-0.0002$ a.u. is larger than the energy errors, the rigorous variational principle from the $N$-representability constraints allows us to obtain energies that are much more accurate than the noise in the quantum computer.}  To the best of our knowledge, these are some of the most accurate energies obtained to date with a generalized quantum architecture. The tolerances of the algorithm are well suited for low iterations on the IBM device, but can be tightened to accomplish lower error rates.

\begin{figure}
\label{fig:polytopes}\begin{center}
\includegraphics[scale=0.57]{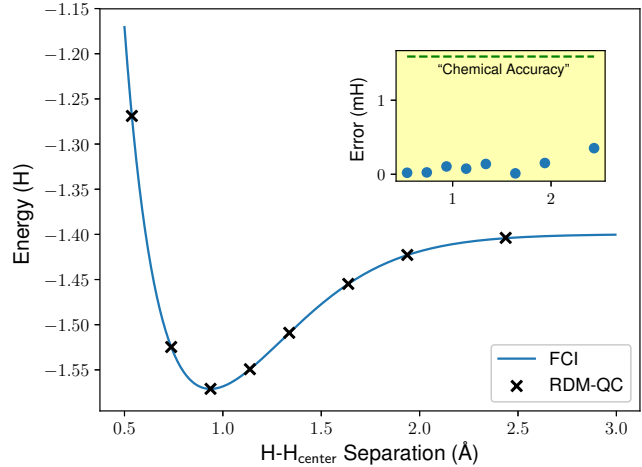}
\caption{Depicts the dissociation curve for the doublet H$_3$ with respect to the bond distance from the center H to the two exterior H atoms in a linear geometry. The crosses were calculated with a variational quantum algorithm on the quantum computer, while the line was generated with a full configuration interaction (FCI) calculation on a classical computer. Energies are listed in Hartrees. The inset plot depicts the \textcolor{black}{error} from the full-CI method as a function of the separated distance, reported in milli-Hartrees, mH. The dashed line is at 1.6 mH, which corresponds to 1.0 kcal/mol, a number that is generally used as a guide for chemical accuracy. \textcolor{black}{The effect of errors is discussed in the text and in Appendix A. Variability from a single run is on the order of $2 - 0.2$ mH, and in our optimization we purposefully oversample the target region.}}
\end{center}
\end{figure}

\subsection{Mott Insulator Transition for H$_3$}

Using the computed 1-RDM, we can also calculate one-electron properties of the system. Upon dissociation molecular H$_{3}$ undergoes a Mott transition from a metal to an insulator. The transition can be observed from sum of the squares of the off-diagonal elements of the 1-RDM in the local atomic-orbital basis set. Figure 2 compares the sum of squares from the RDM calculation on the quantum computer (RDM-QC) with the corresponding results from Hartree-Fock (HF) and FCI. We observe that HF theory fails to capture the metal-to-insulator transition, remaining metallic throughout the dissociation but the RDM-QC correctly predicts the transition in close agreement with FCI. RDM-QC captures this transition because its parameterization captures the requisite strong electron correlation. Figure~1 in the Supplemental Material also shows the curve from second-order many-body perturbation theory which, unlike the algorithm on the quantum computer, breaks down in the region of the bond dissociation.

\begin{figure}\label{fig:polytopes2}
\begin{center}
\includegraphics[scale=0.55]{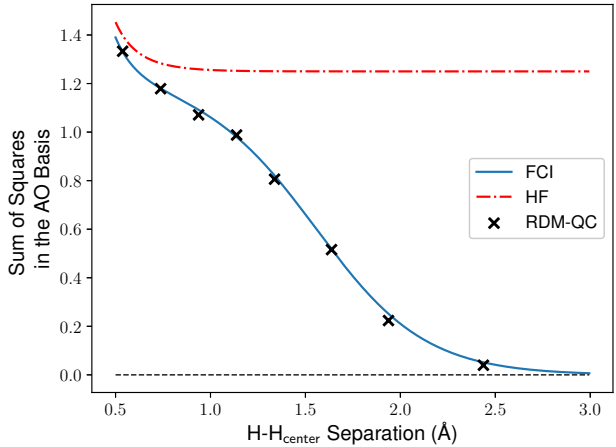}
\caption{Shows the sum of squares of the off-diagonal elements, $\tau$, of the 1-RDM of H$_3$ in the local L{\"o}wdin atomic orbital basis along the dissociation curve of H$_3$. Here, $ \tau = \sum_{i\neq j} \lVert {}^1 D_{j}^{i} \rVert^2$ where $i,j$ are orbital indices in the L{\"o}wdin atomic orbital basis. The Hartree-Fock result is shown as a dashed-dotted line, the FCI result is solid, and our variational quantum computation is shown as crosses. The bottom dashed line shows the dissociated limit where $\tau=0$, and the natural orbitals approach the atomic ones. That H$_3$ serves as a Mott-Insulator can be seen between these distances, as $\tau \rightarrow 0$ with increasing distance, highlighting the mean field and 2-electron approaches.}
\end{center}
\end{figure}

The expression of the pure $N$-representability condition in terms of the natural orbitals, it suggests a natural partition of the electronic structure calculation between the quantum and classical computers. Minimization of the energy as a functional of the 2-RDM in the natural-orbital basis set is performed on the quantum computer while optimization of the natural orbitals is performed by inexpensive, polynomially scaling orbital rotations on the classical computer. In the language of quantum information non-local degrees of freedom, responsible for multi-particle entanglement, are optimized on the quantum computer, and local degrees of freedom are optimized on the classical computer~\cite{Linden1997,Acin2001}. In classical electronic structure the separation of the orbital optimization has precedent in methods like self-consistent-field methods, Brueckner-orbital coupled cluster theory, and natural-orbital functional theories~\cite{Chiles1981,Handy1989,Schmidt1998,Helgaker2000,Piris2007,Mazziotti2007}.
\section{Conclusion}
A quantum-classical hybrid algorithm for molecular electronic structure is implemented that uses a 2-RDM-based energy variational principle in which the energy is minimized with the 2-RDM constrained by $N$-representability conditions. Computations are performed for the strongly correlated dissociation of the H$_{3}$ molecule. The QC-RDM calculation accurately captures the potential energy curve within an error of about 0.1~kcal/mol even in the dissociation region where classical single-reference methods fail. It also yields the 1- and 2-RDMs with the 1-RDM revealing the Mott transition from a metal to an insulator.

While previously employed hybrid algorithms like the variational quantum eigensolver also compute the 2-RDM en route to the energy, the present work uses a 2-RDM-based variational principle in that we explicitly constrain the 1- and 2-RDMs to be pure $N$-representable. The $N$-representability conditions provide a physically motivated \textcolor{black}{error mitigation scheme} which is critical to achieving accurate results on near-term quantum computers which are noisy and prone to errors.  While the present work employs an $N$-representability condition for 3-electron systems due to Borland and Dennis, the 2-RDM-based variational principle on a quantum computer is applicable to systems with arbitrary $N$ through the use of more general $N$-representability conditions. The present work provides an important step towards harnessing two-electron reduced density matrix theory within the context of quantum computing for accurate computations of many-electron molecules and materials.

\section*{Acknowledgements}
\textcolor{black}{The authors acknowledge use of the IBM Q for this work. The views expressed are those of the authors and do not reflect the official policy or position of IBM or the IBM Q team.} D.A.M. gratefully acknowledges the Department of Energy, Office of Basic Energy Sciences, Grant DE-SC0019215, the U.S. National Science Foundation Grant CHE-1565638, and the U.S. Army Research Office (ARO) Grant W911NF-16-1-0152.

\section*{Data Availability}

Data is available from the corresponding author upon reasonable request.

\section*{Author Contributions}

D.A.M. conceived of the research project. S.E.S. and D.A.M. developed the theory.  S.E.S. performed the calculations. S.E.S. and D.A.M. discussed the data and wrote the manuscript.

\appendix

\section*{Appendix A: Computational Details}
The electronic structure package PySCF ~\cite{Sun2017} was used to obtain electron integrals and to perform restricted open-shell Hartree-Fock and full configuration interaction calculations.

For the quantum computation we used the IBM Quantum Experience devices (ibmqx4), available online, with a 5-transmon quantum computing device~\cite{Koch2007}. These cloud accessible quantum devices are fixed-frequency transmon qubits with co-planer waveguide resonators~\cite{Koch2007,Chow2011}. \textcolor{black}{Experimental calibration and connectivity of these devices is included in Appendix D. The quantum information package Qiskit was used for interfacing with the device \cite{Qiskit}.}

\textcolor{black}{A compact qubit mapping was utilized, similar to previous work with a 3-electron in 6-orbital system \cite{Smart2019}, although adapted for the current work. Each evaluation of the quantum computer had 2048 measurements on the 3-qubit populations, with no additional tomography required. About 2 macro iterations were required for most distances (with only one taking 5 iterations). The stopping criteria on the quantum computer was a distance of 0.1$^\circ$ of the simplex vertices to the centroid (although repeatability errors for set parameters were on a larger scale than this). A threshold of 0.1 mH between the quantum computation and the orbital rotations was required for the macro iterations to terminate. We illustrate the optimization with data taken from the point $d=1.34$ in Figure 3.}

\textcolor{black}{Despite the relatively high sampling error on the quantum computer, the algorithm is able to find much higher accuracy answers due to repeated sampling across the valley of the energy surface, as well as ensuring that a system is $N$-representable. Because sampling even within a distance of 2$^\circ$ may lead to highly variable results, the strict criteria of convergence for the simplex implies that the region is well sampled. }

\begin{figure*}
\label{error}
\begin{center}
\includegraphics[scale=0.60]{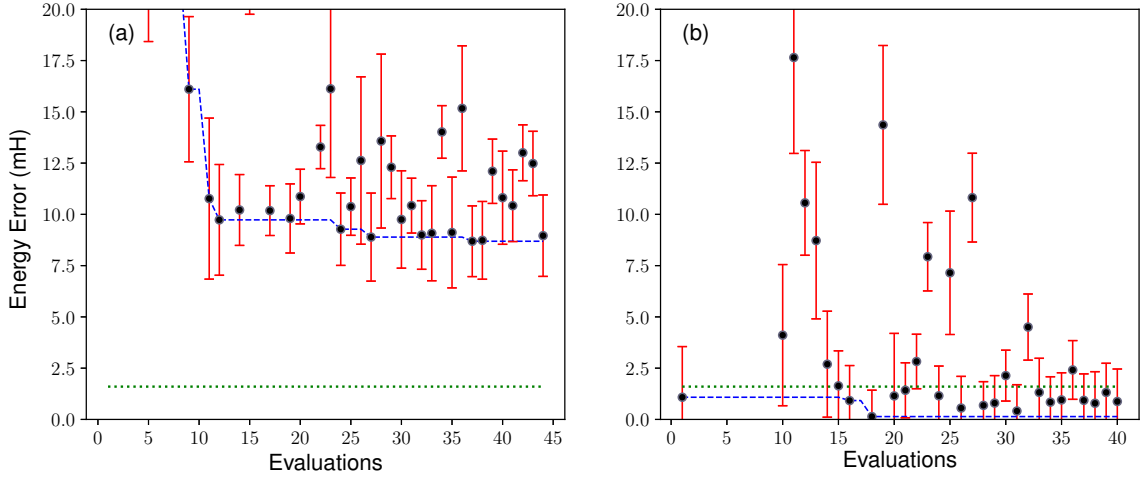}
\caption{\textcolor{black}{The energy errors of the (a) first and (b) second macro iterations of the Nelder-Mead simplex optimization of $H_3$ are shown as a function of the number of energy evaluations.  The variational design of the algorithm allows for lower errors in the final energy than present in the sampling.  The energy errors in milli-Hartrees (mH) are measured relative to the energies from full configuration interaction. The blue dashed line shows the energy of the best simplex point along the optimization, while the green line shows the ``chemical accuracy'' threshold of ~1.6 mH.  Error bars correspond to a 90\% confidence interval based largely on sampling errors in the quantum computer.   The edge hydrogens in H$_{3}$ have a separation of $1.34$ \AA.}}
\end{center}
\end{figure*}

\section*{Appendix B: Mapping the Wavefunction}
\textcolor{black}{Here we explicitly describe the mapping of the wavefunction. We first map the computational qubit states ($q_{ik}$, where $i$ is qubit number, $k$ is the qubit state, 0 or 1) to the GPC orbitals ($n_i$). Then, we use an additional mapping from the GPC orbitals to the molecular spin orbitals ($\phi_i$). The mappings are as follows:
\begin{eqnarray}
q_{1,0} \leftrightarrow n_1 \leftrightarrow \phi_{1\alpha}    \\
q_{2,0} \leftrightarrow n_2 \leftrightarrow \phi_{2\alpha}  \\
q_{3,0} \leftrightarrow n_3 \leftrightarrow \phi_{1\beta}  \\
q_{3,1} \leftrightarrow n_4 \leftrightarrow \phi_{3\alpha}  \\
q_{2,1} \leftrightarrow n_5 \leftrightarrow \phi_{2\beta}  \\
q_{1,1} \leftrightarrow n_6 \leftrightarrow \phi_{3\beta}
\end{eqnarray}
These mappings, albeit useful for the present implementation, are not unique.  The parameters in Appendix D and the preparing gates in Eq.~\eqref{qalg1} account for the selected ordering.}

\section*{Appendix C: Reconstruction of the 2-Electron Reduced-Density-Matrix (2-RDM)}
The wavefunction which is pinned to the Borland-Dennis constraint is given in Eq.~\eqref{gpcwf}. We can also parameterize this in terms of the eigenvalues (occupations) of the 1-electron reduced-density-matrix (1-RDM) $n_i$, and a coefficient phase $p_i$.
\begin{equation}\begin{split}
|\psi \rangle =  p_\alpha \sqrt{1-n_5-n_6} &|111000\rangle \\  + p_\beta \sqrt{n_5}&|100110\rangle \\  + p_\gamma\sqrt{n_6} &|010101\rangle.
\end{split}\end{equation}
The choice of $n_5,n_6$ is somewhat arbitrary, but shows that having only two non-matching $n_i$ is sufficient to represent the wavefunction. From this, the 2-RDM \emph{in the GPC basis} can be constructed as follows. We choose a $S_z=+1/2$ spin state with the following $\alpha / \beta$ orbital assignment: $\alpha \in \{1,2,4\}, \beta \in \{3,5,6\}$. An element of the 2-RDM is given by:
\begin{equation}
^2 D^{i,k}_{j,l} = \langle\psi|\hat{a}^{\dagger}_{i}\hat{a}^{\dagger}_{k}\hat{a}^{}_{l}\hat{a}^{}_{j}|\psi \rangle
\end{equation}
where $\hat{a}_i^\dagger$,$\hat{a}^{}_i$ are the second-quantized creation and annihilation operators, respectively. The $\alpha\alpha$ and $\alpha\beta$ blocks of the 2-RDM are given by:
\begin{equation}
{}^2 D^{\alpha,\alpha}_{\alpha,\alpha} =
\begin{pmatrix}
1-n_5-n_6  & 0 & 0  \\
 0 & n_5 &  0 \\
0 & 0 & n_6 \\
\end{pmatrix},
\end{equation}
with the column basis $\{\hat{a}_2\hat{a}_1,\hat{a}_4\hat{a}_1,\hat{a}_4\hat{a}_2 \}$, used in the $\alpha\alpha$ block,
\begin{table*}
\begin{equation}
{}^2 D^{\alpha,\beta}_{\alpha,\beta} =
\begin{pmatrix}
1-n_5-n_6 & *& *& *& *& *\\
0 & n_5& *& *& *& *& \\
0 & 0&  1-n_5-n_6& *&*& *\\
0 & p_\beta  \sqrt{n_5n_6}& 0&  n_6& *& * \\
0 & 0&  p_\alpha p_\beta \sqrt{(1-n_5-n_6)(n_5)}& 0&   n_5& * \\
-p_\alpha \sqrt{(1-n_5-n_6)(n_6)} & 0& 0& 0& 0&  n_6 \\
\end{pmatrix},
\end{equation}
\end{table*}
and the column basis $\{\hat{a}_3\hat{a}_1,\hat{a}_5\hat{a}_1,\hat{a}_3\hat{a}_2,\hat{a}_6\hat{a}_2,\hat{a}_5\hat{a}_4,\hat{a}_6\hat{a}_5\} $  for the $\alpha\beta$ block (note that $\hat{a}^\dagger_i \hat{a}^\dagger_j (\hat{a}_6\hat{a}_1 / \hat{a}_5\hat{a}_2 / \hat{a}_3\hat{a}_4)$ are all equal to 0 for any $i,j$).  The non-zero elements which require the flexibility of the sign are: $\hat{a}^\dagger_4\hat{a}_6^\dagger\hat{a}^{}_3\hat{a}^{}_1$, $\hat{a}^\dagger_2\hat{a}_6^\dagger\hat{a}^{}_5\hat{a}^{}_1$, and $\hat{a}^\dagger_4\hat{a}_5^\dagger\hat{a}^{}_3\hat{a}^{}_2$ and their Hermitian conjugates ($*$ terms). Additionally, we have set $p_\gamma=1$, but still have equivalent degrees of freedom in the sign of the two remaining terms.  This treatment of the \textcolor{black}{sign terms is included in the next section.}

\section*{Appendix D: Wavefunction Parity Mapping}
In constructing the terms $p_i$, the main requirement is that they do not break the continuity of the potential surface, and that all degrees of freedom are still reachable. Because of the symmetry of the $n_i$ and their repeating structure with respect to the input parameters $\theta_i$, we are able to link $p_i$ with these parameters. We also bound the input $\theta_i$ to the minimal region required to create any point in the plane. Define a new variable $\phi$ which maps $(-\infty,\infty)$ to $[-\pi/4,\pi/4]$:
\begin{equation}
\phi(\theta) = (-1)^{x(\theta)}\Big[(\theta+\frac{\pi}{4})\Mod{\frac{\pi}{2}}\Big]
\end{equation}
where
\begin{equation}
x(\theta) =  \frac{\theta- \frac{\pi}{4}-(\theta+\frac{\pi}{4})\Mod{\frac{\pi}{2}}}{\frac{\pi}{2}}.
\end{equation}
While we will never have switching between the ordering of $n_1$ with $n_3$ or $n_2$ with $n_3$, it is possible to switch between $n_1$ and $n_2$, and our sign mapping should be invariant to this switch. We achieve this invariance by requiring that if $\phi_1<0$, then:
\begin{equation}
\text{ if } \phi_2 \geq -\phi_1 \text{ and } \phi_2 \geq 0 \rightarrow \gamma< 0, \text{ else } \rightarrow \beta<0
\end{equation}
and if $\phi_2<0$:
\begin{equation}
\text{ if } \phi_2 \geq -\phi_1 \text{ and } \phi_2 \geq 0 \rightarrow \gamma< 0, \text{ else } \rightarrow \beta<0.
\end{equation}
Combined, these conditions produce a mapping of the set of $\theta_1,\theta_2$ to a $\pm$ sign, which is symmetric across the line $\theta_1=\theta_2$. This gives the required mapping of signs. While the surface is not smooth because of the boundary around $\frac{n\pi}{4}$, it is continuous.

\section*{{Appendix E: Quantum Computer Errors}}
\textcolor{black}{Calibration data used in obtaining the H$_3$ calculations on the IBM device is included in Table~I\cite{Sparrow2018}.}

\begin{table}[h]\caption{Contains calibration info for the 5 transmon qubit device, denoted as IBMQX Tenerife, Raven, or ibmqx4. Included are the readout errors for the qubits and the single and multi-qubit gate errors for each date. In general, the qubits were selected so as to have minimal error on the device. Qubit connectivity is directional with the first qubit indicating the control and second qubit indicating the target.}
\begin{center}

\begin{tabular}{cccccccc }
\hline
\textbf{Calibration Date:} & 1& 2 & 3 & 4 & 5 & 6\\
\hline
\multicolumn{7}{l}{\textbf{Readout Error ($10^{-3}$)}}  \\
\hline
Q0: & 64 & 74 & 66 & 56 & 66 & 87\\
Q1: & 65 & 75 & 52 & 43 & 43 & 75\\
Q2: & 16 & 19 & 18 & 18 & 25 & 19\\
Q3: & 38 & 13 & 24 & 33 & 30 & 81\\
Q4: & 45 & 37 & 208& 219& 323 & 309\\
\hline
\textbf{Gate Error ($10^{-3}$)} \\
\hline
Q0: & 0.77 & 1.03 & 1.20 & 0.86 & 0.69 & 0.86\\
Q1: & 1.80 & 2.32 & 1.63 & 1.72 & 1.29 & 1.20\\
Q2: & 1.03 & 1.20 & 0.94 & 1.03 & 1.46 & 0.86\\
Q3: & 1.46 & 1.63 & 1.29 & 1.63 & 1.80 & 1.72\\
Q4: & 1.37 & 1.29 & 1.63 & 1.20 & 3.35 & 3.35\\
Q1$-$Q0: & 29.2 & 37.5 & 32.9 & 24.6 & 27.9 & 32.5\\
Q2$-$Q0: & 30.5 & 27.5 & 28.6 & 24.9 & 31.2 & 32.6\\
Q2$-$Q1: & 29.9 & 42.7 & 34.5 & 32.7 & 41.4 & 34.5\\
Q3$-$Q2: & 43.3 & 69.6 & 50.8 & 65.7 & 60.6 & 59.3\\
Q3$-$Q4: & 36.9 & 45.1 & 40.8 & 36.6 & 69.9 & 79.5\\
Q4$-$Q2: & 43.8 & 46.3 & 55.5 & 47.8 & 91.0 & 72.2\\
\end{tabular}
\end{center}

\end{table}
\pagebreak

\bibliography{qc_chem}
\bibliographystyle{apsrev4-1}

\end{document}